# Indistinguishable photons from a diode

A. J. Bennett[1], R. B. Patel[1, 2], A. J. Shields[1], K. Cooper[2], P. Atkinson[2,a], C. A. Nicoll[2] and D. A. Ritchie[2]

[1] *Toshiba Research Europe Limited, Cambridge Research Laboratory, 208 Science Park, Milton Road, Cambridge, CB4 OGZ, U. K.*

[2] *Cavendish Laboratory, Cambridge University, JJ Thomson Avenue, Cambridge, CB3 0HE, U K.*

[a] *Present address Max-Planck-Institut für Festkörperforschung, Heisenbergstr.1, D-70569 Stuttgart, Germany*

## Abstract

We generate indistinguishable photons from a semiconductor diode containing a InAs/GaAs quantum dot. Using an all-electrical technique to populate and control a single-photon emitting state we filter-out dephasing by Stark-shifting the emission energy on timescales below the dephasing time of the state. Mixing consecutive photons on a beam-splitter we observe two-photon interference with a visibility of 64%.

## Main text

If two separate identical photons impinge on a beam-splitter from opposite directions at the same time they appear to leave in the same direction[1]. This provides a route to generating entanglement, either between the photons themselves[2,3] or the sources[4]. The original demonstrations of two-photon interference were made with optically-pumped non-linear crystals generating pairs of photons via parametric down-conversion[1,2]. When operated at a low pump-level with appropriate filtering this technique can produce high visibility interference. Nevertheless, the need for single-photon sources which can generate indistinguishable photons has stimulated research on fluorescence of single atoms and ions in a cavity[5,6], single molecules[7], atomic ensembles[8] and single semiconductor quantum dots (QDs) in cavities[9]. In the case of solid-state systems in particular, the interference visibility is reduced by interactions between the state and surrounding electronic and vibronic states which destroys the phase and energy relationship between the successive photons. Indistinguishability can be improved by selective resonant optical excitation of the state only and coupling the emitter to a high-quality factor, low-volume cavity to decrease the spontaneous emission rate to the point where dephasing is limited by the radiative emission[9].

Here we demonstrate a different route to eliminate the effects of dephasing that does not require quasi-resonant excitation or coupling to a high-quality, low-volume cavity. We use a miniature, fast and robust single-photon-emitting-diode[10,11] based on a single QD in a p-i-n diode. We report a new mode of operation that reduces the probability of multi-photon emission significantly and increases the indistinguishability of the photons. This is achieved by selecting a narrow range of energies for photon collection thereby reducing the range of emission times to 31ps, well below the dephasing time of the state.

The experimental arrangement used to measure the degree of indistinguishability of photons from the diode is shown in Fig. 1(a). Two photons separated by a time difference $\delta\tau_0$ are coupled into a Mach-Zehnder interferometer in polarization-maintaining single-mode fibre where one arm is longer by $\delta\tau_1$. There is then only one way that the two photons can be coincident at the beam-splitter $C_B$: when the 1st photon takes the longer path (where it is delayed by $\delta\tau_1$) and the 2nd photon takes the shorter path. When this event occurs, if the photons are indistinguishable in polarisation and energy and $\delta\tau_0 = \delta\tau_1$, then the two detectors $D_1$ and $D_2$ will observe a reduction in coincident detections. Distinguishability can be restored either by ensuring the photons have opposite polarization at $C_B$, or by changing $\delta\tau_0$ such that $\delta\tau_0 \neq \delta\tau_1$.

The sample we use to perform our experiment is a planar micro-cavity p-i-n diode with 2 (12) periods of Bragg mirror above (below) a single layer of low-density InGaAs/GaAs self assembled QDs at the centre of a $\lambda$ spacer layer[11,12]. We use a negatively charged exciton state emitting at 1.31495 eV (marked with an arrow in Fig. 1 (b)). It is important to state here that the cavity is employed to increase the efficiency with which photons are collected[11]: it has no measurable effect on the spontaneous emission rate of the quantum dot. During the experiment the diode is cooled to 4K and emission is collected normal to the surface by a microscope objective with NA=0.5. A polarizing beam-splitter and monochromator are used to filter the photons to have a well defined linear polarization and energy, before coupling into the interferometer. Photons are detected with two silicon avalanche-photo-diodes ($D_1$ and $D_2$), connected to a photon-counting card.

For optimum performance of the diode we apply the voltage signal shown in Fig. 2(a) which Stark shifts the energy of the state (Fig. 2(b)). For most of each emission cycle the DC bias is ~1.45V so no current flows. Short (nominally 300ps) positive-amplitude pulses are used to inject carriers into the diode. 300 ps later a negative-amplitude voltage pulse is used to shift the emission line into the spectral window where the photon is collected at 1.31475eV, leading to the time-resolved photon detection trace shown in Fig. 2(c). By reducing the length of this negative-amplitude electrical pulse the jitter in the time at which the photons are transmitted to the detectors is controlled. During the time in which the photon is collected there is no current flowing through the device, reducing dephasing. In addition, this pulse sequence eliminates refilling of the state within each cycle and the negative-amplitude pulse preferentially depopulates other states in the diode, reducing the level of background emission, both of which could lead to an increased proportion of multi-photon emission. A typical Hanbury-Brown and Twiss[13] measurement is shown in Fig. 3(a) where $g^{(2)}(0) = 0.03 \pm 0.01$, which is comparable with the lowest values reported from similar dots under quasi-resonant excitation[9]. We do not observe any bunching or anti-bunching of photons in adjacent pulses.

Using the same state and operating conditions we then carried out two-photon interference experiments with a driving voltage repetition period of $\delta\tau_0 = 1.98$ ns, matched to the interferometer delay $\delta\tau_1 = 1.98$ ns. When the experiment is performed with two photons of parallel polarization meeting at the final coupler $C_B$ there is a suppression of the peak at time zero (Fig. 3(b)). As a control experiment, measurements were taken for the case where the photons have orthogonal polarizations at $C_B$. In this case the area of the time-zero peak is, within error, half that of the peaks at time > 4ns, as expected (Fig. 3(c)). When the photons are distinguishable the peak at time zero should have a relative area of 0.5, the peaks at $\pm\delta\tau_0$ should be at 0.75 and all other peaks unity. If the photons are indistinguishable the area of the central peak is reduced. The areas stated in Fig. 3(d) are normalised such that the mean of the ten peaks at $\pm 2\delta\tau_0$, $\pm 3\ \delta\tau_0$, $\pm 4\ \delta\tau_0$, $\pm 5\ \delta\tau_0$ and $\pm 6\ \delta\tau_0$ is equal to unity.

To further characterise the source a series of measurements were taken with variable source repetition periods and we observed a characteristic "dip" in the coincidence rate as shown in Fig. 3(d). From this we are able to determine an interference visibility of 60 ± 4% in the raw data. Subtracting the effect of dark counts in the detectors, the interference visibility for photons from the source is inferred to be 64 ± 4% which for balanced beam-splitters is equal to the square of the photon's wave-function overlap, $<\Psi_1|\Psi_2>^2$ [14]. Increased visibility could be achieved by selecting a quantum dot state with a longer coherence time or reducing the temporal length of the pulse further. A simple criterion for whether a single photon source can be used to generate polarization entanglement in the scheme of Fattal *et al*[14] is that the visibility is greater than $2g^{(2)}(0)$. This criterion is fulfilled by our source.

Finally, we would like to consider the form of the dip in the coincidence detection probability. To do this we have built upon the model of Legero *et al*[16], reformulating their theory to cover photons with a Lorentzian spectrum. We include in this model a Gaussian timing jitter on the photon emission time which has the effect of reducing the visibility of the dip. We also include the effect of the time varying Stark shift but surprisingly we find that this has negligible effect on the shape, width or depth of the dip shown in Fig. 3(d) for the parameters used in our experiment. We find a good fit to the measured dip if we assume there is a Gaussian jitter on the time at which the centre of each photon wave-packet is emitted, with width of 31ps, and that each photon has a 1/$e$ coherence time of 60ps. Independent measurements of the photon coherence time yield a similar value under pulsed operation. Using the standard formula, $1/T_2 = 1/2T_1 + 1/T_2^*$[7], and the radiative lifetime, $T_1$ = 800ps we estimate the dephasing time $T_2^*$ = 62ps. We note that the visibility of interference that would be achieved if optically exiting this QD under fixed bias is given by the well-known formula Visibility ~ $T_2/2T_1$[15] = 3.75%, well below the value we measure here using this new technique. In conclusion, in our experiment we see an increased visibility of two-photon interference because we are selecting photons emitted in a narrow time range which is less than the dephasing time, thus we are filtering out the temporal information which may distinguish photons.

## Fig 1

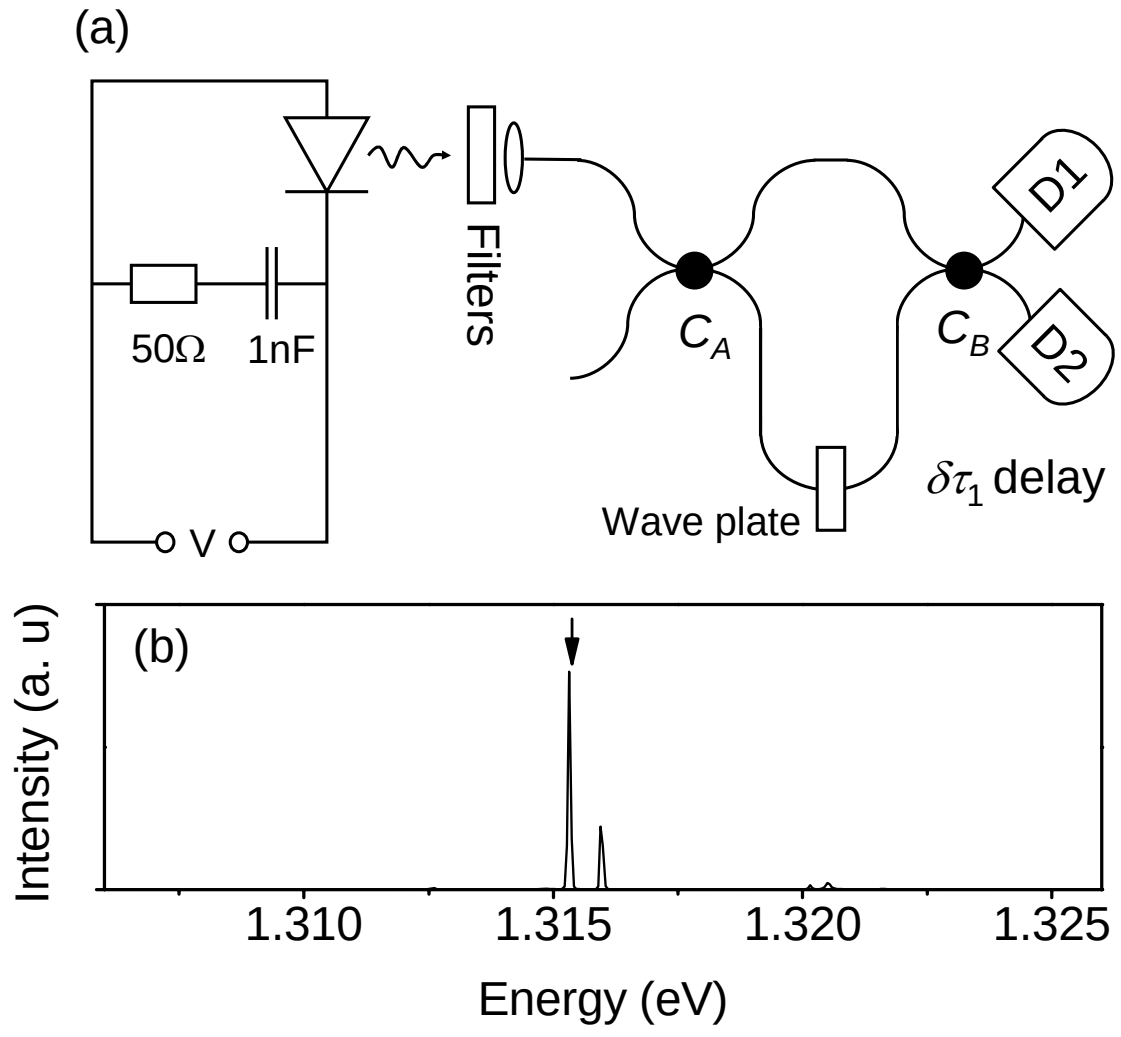

## Fig. 2

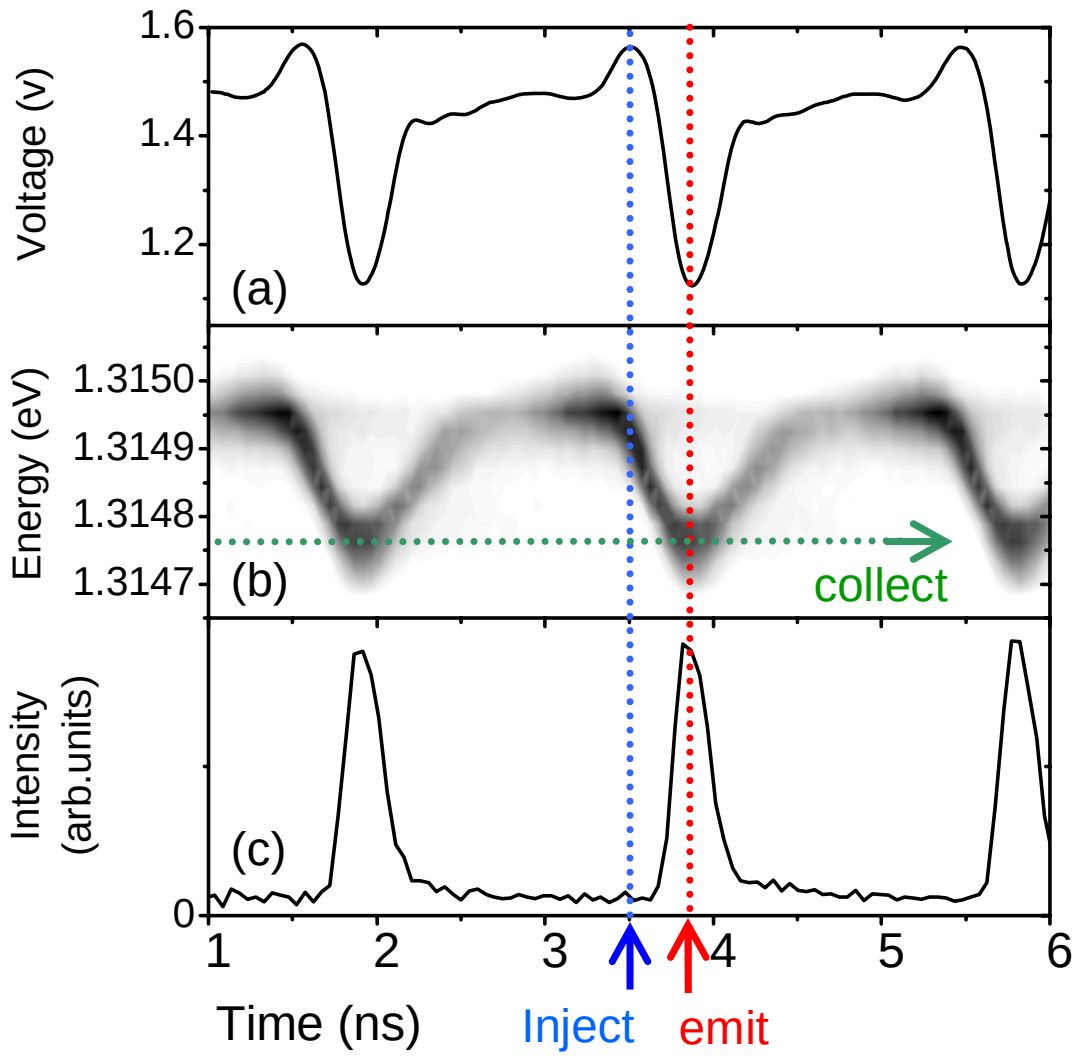

## Fig. 3

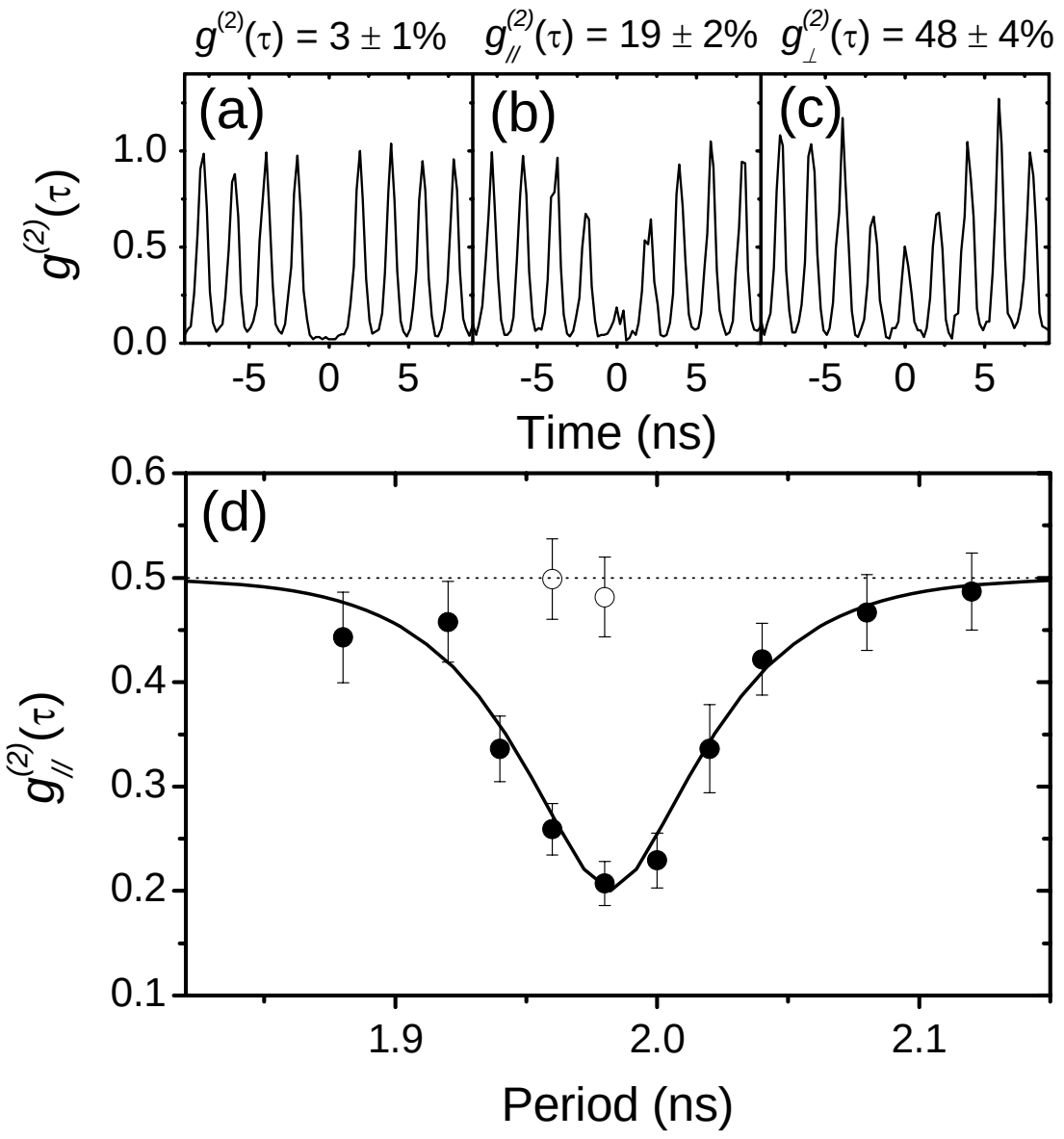

## Figure Captions

Fig. 1. (a) Schematic layout of our experiment. On the left hand side a circuit is used to drive the diode. Photons are coupled into the unbalanced fibre-optic Mach-Zehnder interferometer where the two photons are able to interfere on the final coupler $C_B$. (b) Emission spectrum under DC bias of 1.45V and current of 100μA.

Fig. 2. (a) The voltage sequence employed. A positive height pulse injects carriers to the quantum dot and the voltage is then immediately reduced by 0.61V to allow a photon to be emitted at the desired energy. (b) Emission of the quantum state showing the time-varying Stark shift due to the applied voltage pulse. (c) If photons are collected only when the voltage is minimal the source delivers photons at well-defined times.

Fig. 3. (a) Hanbury-Brown and Twiss correlation showing a 30-fold suppression of multi-photon emission. (b) Two-photon interference correlation recorded with parallel single photons and (c) with orthogonal photons. The relative area of the central peak when performing interference with parallel photons (filled symbols) as a function of the driving voltage repetition rate is shown in (d). A clear "dip" with 60 % visibility is observed. When the photons' polarization is orthogonal no interference is observed (open symbols).

**Acknowledgements** This work was partially funded by the EPSRC and EU-projects QAP and SANDiE.